\documentclass[numberedappendix]{emulateapj}
\usepackage{amssymb,amsmath}
\usepackage{verbatim}
\usepackage{color,hyperref}
\usepackage{needspace}
\usepackage{enumitem}
\usepackage{tikz}
\usepackage{graphicx}
\usetikzlibrary{positioning}
\definecolor{linkcolor}{rgb}{0,0,0.25}
\hypersetup{
  colorlinks=true,        
  linkcolor=linkcolor,    
  citecolor=linkcolor,    
  filecolor=linkcolor,    
  urlcolor=linkcolor      
}
\newcounter{address}
\newcommand{\etal}{et al.}

\newcommand{\eg}{e.g.}

\newcommand{\kpc}{\ensuremath{\,\mathrm{kpc}}}
\newcommand{\pc}{\ensuremath{\,\mathrm{pc}}}
\newcommand{\kms}{\ensuremath{\,\mathrm{km\ s}^{-1}}}

\newcommand{\inv}{\ensuremath{^{-1}}}

\shorttitle{A lack of $L_z=0$ stars in the Solar neighbourhood}
\shortauthors{Hunt, Bovy \& Carlberg}

\submitted{}

\begin{document}

\title{Detection of a dearth of stars with zero angular momentum in
  the solar neighbourhood}

\author{Jason A.~S.~Hunt\altaffilmark{1}, Jo~Bovy\altaffilmark{1,2,3}, and Raymond~G.~Carlberg\altaffilmark{2}}
\altaffiltext{\theaddress}{\label{1}\stepcounter{address} 
  Dunlap Institute for Astronomy and Astrophysics, University of Toronto, 50 St. George Street, Toronto, Ontario, M5S 3H4, Canada}
\altaffiltext{\theaddress}{\label{2}\stepcounter{address} 
  Department of Astronomy and Astrophysics, University of Toronto, 50
  St.  George Street, Toronto, ON, M5S 3H4, Canada}
\altaffiltext{\theaddress}{\label{3}\stepcounter{address} 
  Alfred~P.~Sloan~Fellow}

\begin{abstract}
We report on the detection in the combined \emph{Gaia}-DR1/RAVE data
of a lack of disk stars in the solar neighbourhood with velocities
close to zero angular momentum. We propose that this may be caused by
the scattering of stars with very low angular momentum onto chaotic,
halo-type orbits when they pass through the Galactic nucleus. We model
the effect in a Milky-Way like potential and fit the resulting model
directly to the data, finding a likelihood ($\sim2.7\sigma$) of a dip
in the distribution. Using this effect, we can make a
dynamical measurement of the Solar rotation velocity around the
Galactic center:
$v_{\odot}=239\pm9\kms$. Combined
with the measured proper motion of Sgr A$^*$, this measurement gives a measurement of the distance to the Galactic centre: $R_0=7.9\pm0.3\kpc$.
\end{abstract}

\keywords{
  Galaxy: disk
  ---
  Galaxy: fundamental parameters
  ---
  Galaxy: kinematics and dynamics
  ---
  Galaxy: nucleus
  ---
  solar neighborhood
  ---
  stars: kinematics and dynamics
}

\section{Introduction}\label{sec:intro}
We are entering a golden age for Milky Way astronomy. The European
Space Agency (ESA)'s $Gaia$ mission \citep{GaiaMission}, which
launched $19^{th}$ December 2013, has recently published its first
data release \citep{GaiaDR1}, giving us a new window on our Galaxy,
and in particular, the solar neighbourhood. The primary astrometric
catalogue in \emph{Gaia}-DR1 is the Tycho-$Gaia$ Astrometric Solution
\citep[TGAS,][]{M15,Lindegren16a}, which uses data from the $Tycho$-2
catalogue \citep{H00}, to provide a baseline of approximately 30 years
upon which to calculate astrometric values for stars in common between
$Tycho$-2 and $Gaia$. There is also significant overlap between stars
in the TGAS catalogue and stars observed by the Radial Velocity
Experiment \citep[RAVE, e.g.][]{Rave}, enabling the full 6-dimensional
phase space information to be known for over 200,000 stars in the
solar neighbourhood. This enables us to explore local dynamics in
unprecedented detail.

Many aspects of the structure and dynamics of the Milky Way are
difficult to measure, owing to our position within the Galaxy, and
complex observational selection effects such as dust extinction. Thus,
many of the fundamental parameters of the Milky Way carry significant
uncertainty. One such parameter is the velocity at which the Sun
rotates around the Galaxy, for which plausible values span the range
from $\approx240\kms$ to $260\kms$ \citep[\eg,][]{Bovy12a,Reid14a} and
are dependent on the data set and technique used.

\cite{CI87} proposed that the rotation velocity of the Sun may be
measured by searching for a lack of stars exhibiting zero angular
momentum\footnote{\citet{CI87} phrased this as a measurement of the
  circular velocity. Because the measurement is based on stellar
  velocities with respect to the Sun, the quantity that is directly
  measured is the Sun's motion around the center, not the circular
  velocity. The difference between these two is still quite uncertain
  \citep[\eg,][]{Schoenrich10a,Bovy15b}}. Stars with zero angular
momenta are expected to plunge into the Galactic nucleus and
subsequently experience scattering onto chaotic orbits with a high
scale height, henceforth spending the majority of their time in the
stellar halo \citep{Martinet}. If stars with very low angular momentum
are indeed not present in the solar neighbourhood, the tail of the
tangential velocity distribution will exhibit a dip centred at the
Solar reflex value. This method for measuring the Solar velocity is
attractive because it should depend only on the existence of low
angular momentum orbits within the Milky Way's disk. With
6-dimensional phase space measurements for nearby stars it becomes
possible to calculate the tangential velocity distribution with
respect to the Sun, and thus we can test this prediction by searching
for a dearth of stars around the assumed value of the negative of the
Solar motion.

This paper is constructed as follows. In Section \ref{sec:observation} we discuss our treatment of the data, and the feature observed in the resulting velocity distribution. In Section \ref{sec:modeling} we present simulated models which can explain the feature and make predictions for the size and shape of the observed feature. In Section \ref{sec:detection} we fit our model to the data and present our measurement of $v_{\odot,\rm{reflex}}$. Finally, in Section \ref{sec:conclusion} we discuss the implications of the detection and look forward to future measurements.

\section{Observed feature}\label{sec:observation}
We cross match the TGAS catalogue \citep{Lindegren16a} and the RAVE DR5 data \citep{Kunder16a} to add RAVE line-of-sight velocities to the TGAS astrometric data. Where available, we employ the RAVE spectrophotometric distance estimates, because estimating distances from the TGAS parallax is non-trivial \citep[e.g.][]{BJ}. However, where no RAVE distance is available, we naively invert the TGAS parallax $\pi$ to obtain distance estimates for the remaining stars, removing any star with $\sigma_{\pi}/\pi>0.1$ to avoid large distance uncertainties. The resulting sample consists of 216,201 stars with 6-dimensional phase-space measurements. Then, we convert the velocities from equatorial coordinates to standard Galactic Cartesian coordinates centred on the Sun, $(X,Y,Z,v_X,v_Y,v_Z)$, with $v_X$ positive in the direction of the Galactic centre (toward Galactic longitude $l = 0$) and $v_Y$ positive in the direction of Galactic rotation (toward $l=90^\circ$), both measured with respect to the Sun.

\begin{figure}
\begin{tikzpicture}
  \node (img)  {\includegraphics[width=0.49\textwidth]{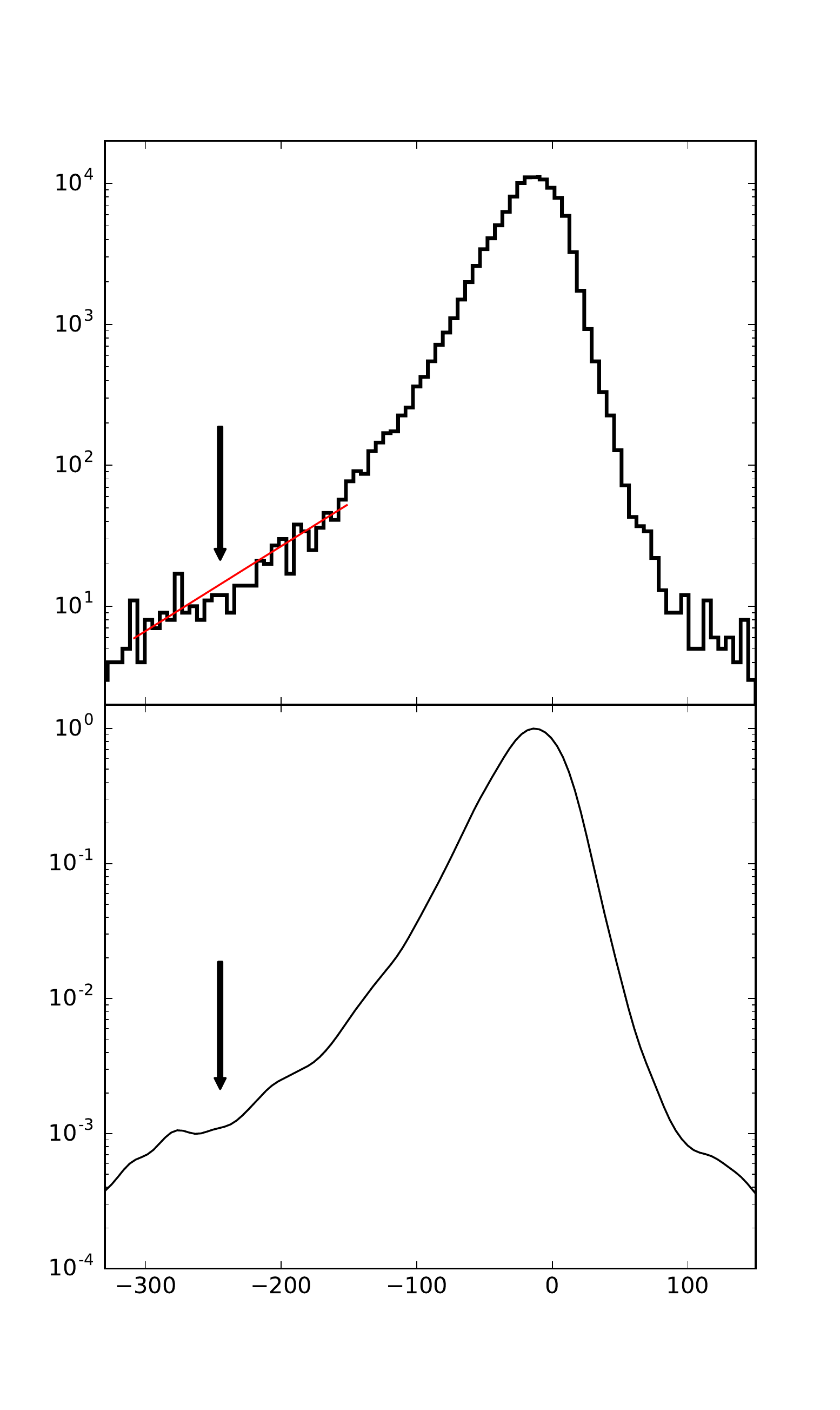}};
  \node[below=of img, node distance=0cm, yshift=2.0cm,font=\color{black}] {$v_Y$ (km s$^{-1}$)};
  \node[left=of img, node distance=0cm, rotate=90, anchor=center,yshift=-1.3cm,xshift=3cm,font=\color{black}] {Number of stars per bin};
  \node[left=of img, node distance=0cm, rotate=90, anchor=center,yshift=-1.3cm,xshift=-3cm,font=\color{black}] {PDF};
\end{tikzpicture}
\caption{\textbf{Top panel:} Distribution of $v_Y$ for stars observed to be within 700 pc of the Sun. The Sun's velocity has $v_Y = 0$. The low-velocity tail of the distribution displays a clear dip in the range between $-210\kms$ and $-270\kms$ that is marked with an arrow and overlaid with an exponential for illustration. \textbf{Bottom panel:} Normalised KDE of the above distribution clearly showing the dip.}
\label{fig:disp}
\end{figure}
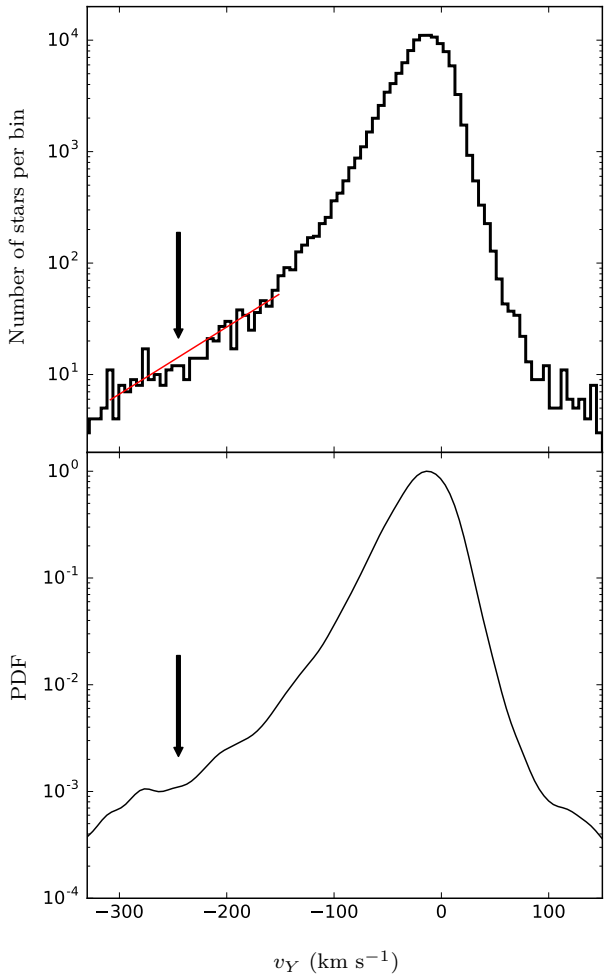

Figure~\ref{fig:disp} shows the distribution of $v_Y$ for stars within
700 pc as a histogram (top panel) and with Kernel Density Estimation (e.g., \citealt{KDE}; bottom panel) using a Gaussian kernel with bandwidth 9. We chose 700 pc as a balance between quantity of stars, and quality of data. There is a clear underdensity of stars in the approximate region of $-210\kms$ to $-270\kms$, marked with an arrow (see also the
top panel of Figure~\ref{fig:fit}). Although there are only
$\approx\!10$ stars per bin at these velocities, this dip in the
distribution is clear across $\approx\!8$ bins. We defer a discussion of the significance of the dip until Section 4 where we fit a model for the dip obtained from simulations.

Furthermore, in the two-dimensional distribution of $(v_X,v_Y)$, we
find a lack of stars with both positive $v_X$ and large negative
$v_Y$. That is, there are very few stars observed on disk orbits
($|v_X| < 150\ \kms$ and $|v_Z| < 150\kms$) plunging towards the
Galactic centre. Quantitatively, there are 24 stars with $v_X > 0$ in
the range $-290\kms < v_Y < -230\kms$ and 34 stars with $v_X <
0$. These rates are inconsistent with being drawn from the same
distribution at $\gtrsim1.7\sigma$.

Our analysis does not explicitly take into account the uncertainties on the distance or velocity estimates. Propagating the uncertainties through the coordinate transformation results in an uncertainty of approximately 10 per cent on the measured cartesian velocities. At the $v_Y$ range of the dip, this corresponds to uncertainties of approximately $24\kms$. Because the width of the dip feature appears to be approximately $60 \kms$ we can neglect the uncertainties in this initial investigation.

\section{Expectation from Galactic models}\label{sec:modeling}

A likely explanation for these missing stars, as mentioned in
Section~\ref{sec:intro}, is that they have been scattered onto chaotic
orbits with larger scale heights by interaction with the Galactic
nucleus. This is expected for disk stars with approximately zero
angular momentum as discussed in \citet{CI87}. Thus, as these stars
would then spend the majority of their orbits far from the Galactic
plane, it is very unlikely that they would be observed in the solar
neighbourhood at any one given time. In Galactocentric coordinates,
such stars have tangential velocities $v_T \approx 0$, corresponding
to heliocentric $v_Y \approx v_{\odot,\rm{reflex}}$, minus the Solar
tangential velocity measured in the Galactocentric frame.

\cite{CI87} performed an analysis of this effect and matched models of the dip constructed within an analytic potential to data from local stellar catalogues complete to 25 pc. They found the dip to be centred at $250\kms$, with a depth greater than 80\,\%. However, they are careful to note that there are only 18 stars with $v_Y<-140\kms$ and that a larger sample will improve the measurement. The TGAS+RAVE sample used here has 374 stars with $-310\kms\leq v_Y<-150\kms$, allowing far greater confidence in our subsequent analysis of the feature.

Firstly, we make a fresh prediction of the feature which we expect to observe in the TGAS data, similar to \cite{CI87}, but with an updated potential, and drawing the distribution of initial positions, radial, and vertical velocities from the observed data to tailor the prediction to the current data set. It is challenging to construct an $N$-body model with sufficient resolution to predict the high velocity tail in the local neighbourhood. Thus, we integrate test particles in a Milky-Way like potential, and observe the resulting orbits.

For our Milky-Way potential we use \texttt{MWPotential2014} from
\texttt{galpy } \citep{Bovy15a}. \texttt{MWPotential2014} consists of
a power-law spherical bulge potential with an exponential cut-off, a
Miyamoto-Nagai disk potential, and a NFW halo potential. The
parameters of this potential have been fit to a wide variety of
dynamical data in the Milky Way; the full parameters are given in
\citet{Bovy15a}. To model the hard Galactic nucleus that is not
included in \texttt{MWPotential2014}, we also include a Plummer
potential $\Phi(R,z)=-M/\sqrt{R^2+z^2+b^2}$,
where $M=2\times10^9$ $M_{\odot}$ and $b=250$ pc. Note that we do
not include a non-axisymmetric bar potential in this initial work,
which may affect the feature slightly. This should be considered when
applying this technique to future $Gaia$ data releases.

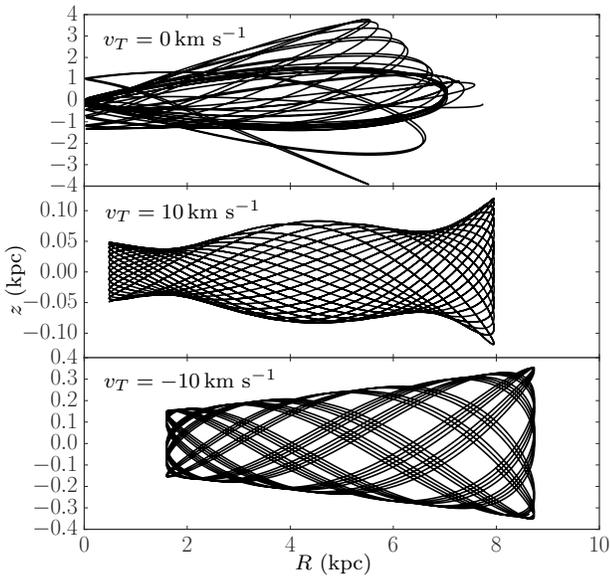
\begin{figure}
\begin{tikzpicture}
  \node (img)  {\includegraphics[width=0.49\textwidth]{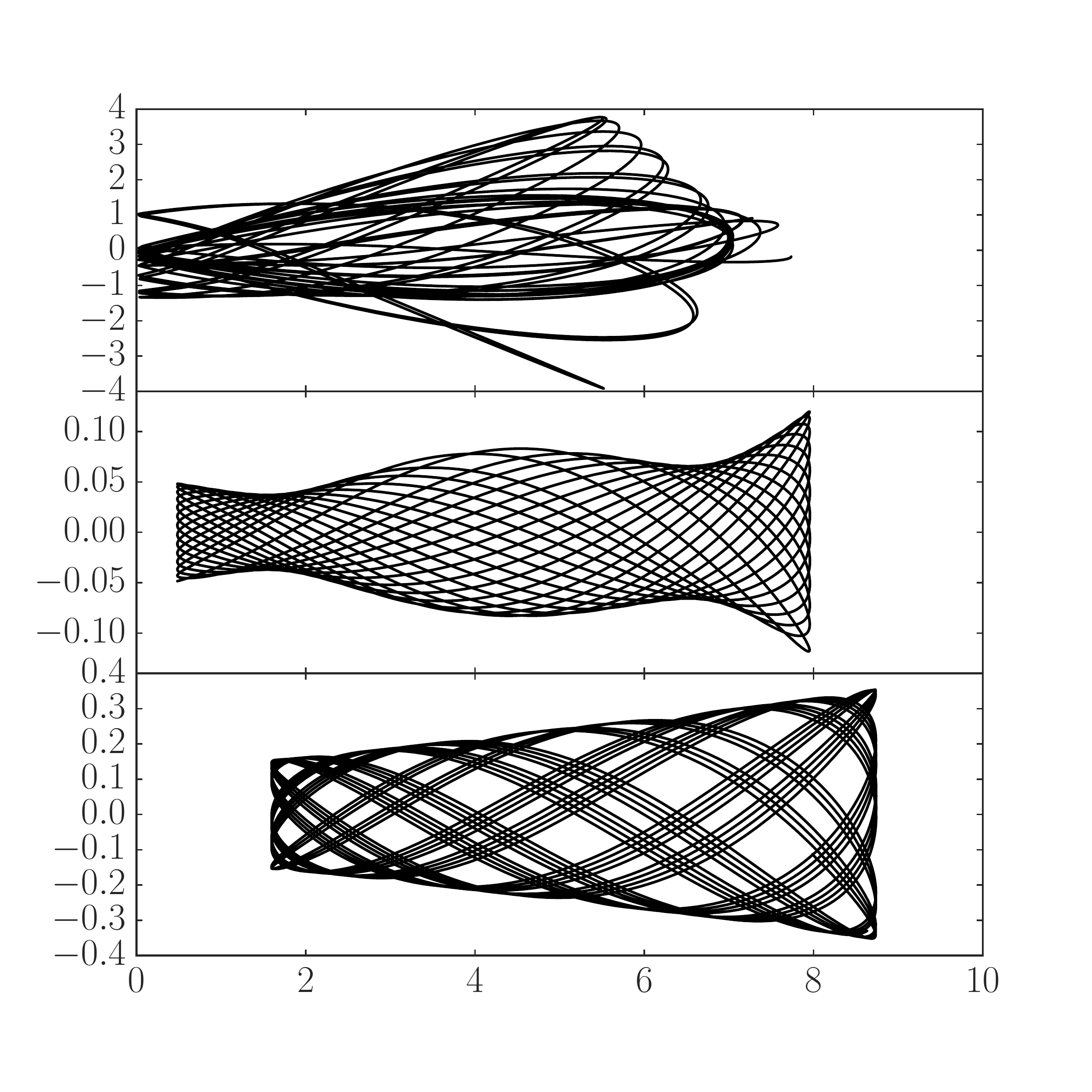}};
  \node[below=of img, node distance=0cm, yshift=2cm,font=\color{black}] {$R$ (kpc)};
  \node[below=of img, node distance=0cm, yshift=9cm,xshift=-2.1cm,font=\color{black}] {$v_T=0\kms$};
  \node[below=of img, node distance=0cm, yshift=4.45cm,xshift=-1.9cm,font=\color{black}] {$v_T=-10\kms$};
  \node[below=of img, node distance=0cm, yshift=6.7cm,xshift=-2.0cm,font=\color{black}] {$v_T=10\kms$};
  \node[left=of img, node distance=0cm, rotate=90, anchor=center,yshift=-1.3cm,,font=\color{black}] {$z$ (kpc)};
\end{tikzpicture}
\caption{\textbf{Top panel:} Example orbit of a particle with $v_T=0\kms$ exhibiting chaotic behaviour. \textbf{Middle panel:} Example orbit of a particle with $v_T=10\kms$. \textbf{Bottom panel:} Example orbit of a particle with $v_T=-10\kms$. Orbits that penetrate the galactic nucleus are scattered onto non-disk orbits that spend little time in the solar neighborhood.}
\label{fig:orbits}
\end{figure}

Figure \ref{fig:orbits} displays the orbits of three stars integrated for 2 Gyr. The top panel shows the orbit of a star with $v_T=0\kms$, which exhibits chaotic behaviour upon interaction with the galactic nucleus. The middle and lower panels show the orbits of stars with $v_T=10\kms$ and $v_T=-10\kms$, respectively, which exhibit well-behaved disk orbits. The star with no angular momentum spends little of its orbital period near the galactic plane, whereas the two stars which do not approach the nucleus remain within a few hundred parsecs of the plane. 

Without an ab-initio model for the disc, it is difficult to predict
how many stars are expected to be missing near zero angular momentum
and what the exact profile of the dip should be. Therefore, we use the
fraction of the orbital period of the test-particle stars that they
spend near the mid-plane of the disc as a proxy for whether it has
been scattered to a much higher scale height. We integrate many orbits
with positions, radial, and vertical velocities drawn from the
TGAS+RAVE sample and from an initial, uniform distribution in $v_Y$
covering the low-velocity tail. We then re-weight the stars using the
fraction of time they spend near the plane and construct the dip
profile by dividing this weighted, final $v_Y$ distribution by the
initial, uniform distribution.

\begin{figure}
\begin{tikzpicture}
  \node (img)  {\includegraphics[width=0.47\textwidth]{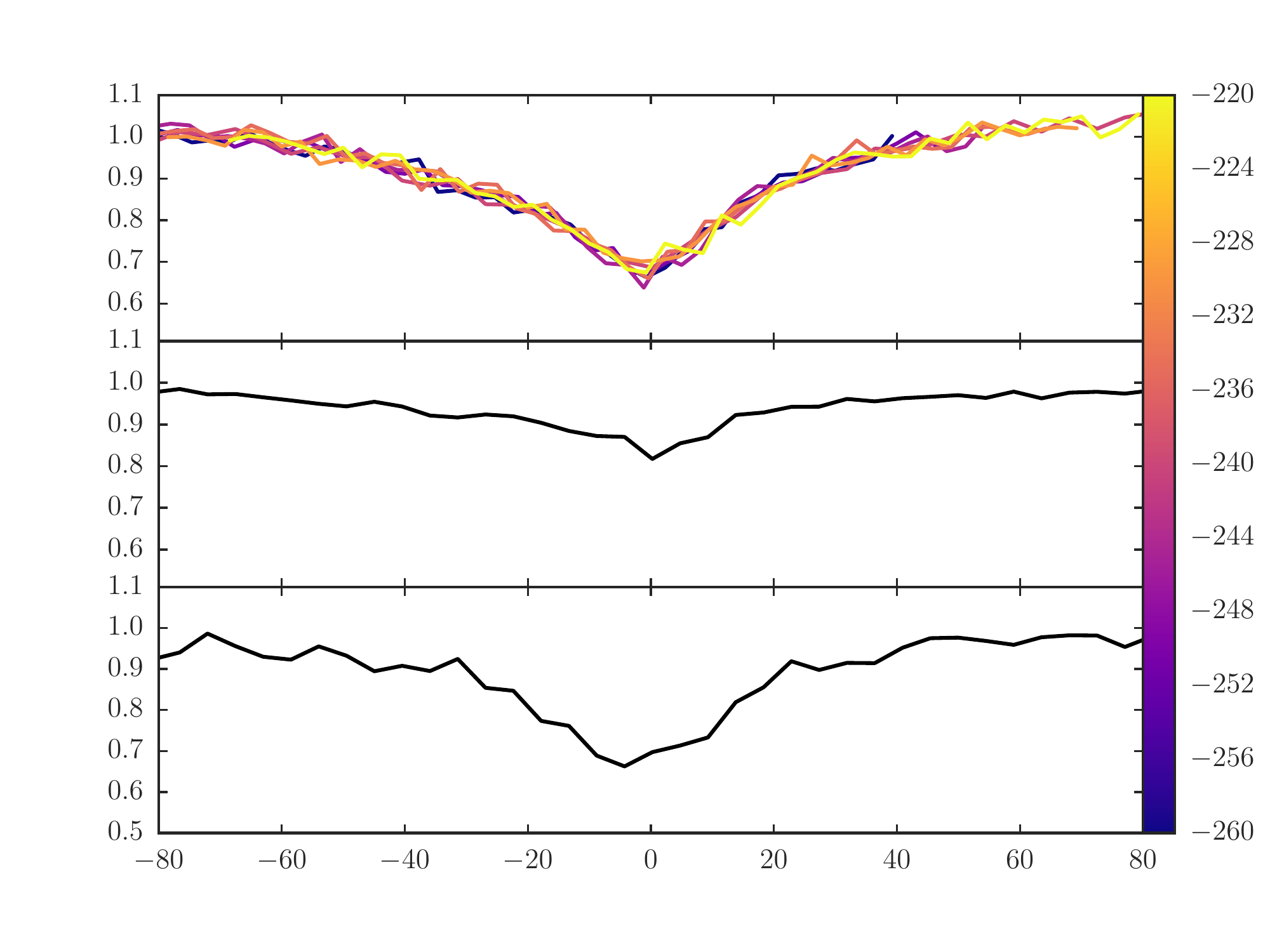}};
  \node[below=of img, node distance=0cm, yshift=1.6cm,font=\color{black}] {$v_Y-v_{\odot,\mathrm{reflex}}$ (km s$^{-1}$)};
  \node[below=of img, node distance=0cm, yshift=5.9cm,xshift=-2.13cm,font=\color{black}] {$|z|<300$ pc};
  \node[below=of img, node distance=0cm, yshift=2.5cm,xshift=-2.2cm,font=\color{black}] {$|z|<50$ pc};
  \node[below=of img, node distance=0cm, yshift=4.2cm,xshift=-2.2cm,font=\color{black}] {$|z|<1$ kpc};
  \node[left=of img, node distance=0cm, rotate=90, anchor=center,yshift=-1.5cm,,font=\color{black}] {Fraction of stars remaining};
\node[right=of img, node distance=0cm, rotate=270, anchor=center,yshift=-1.0cm,,font=\color{black}] {$v_{\odot,\mathrm{reflex}}$ (km s$^{-1})$};
\end{tikzpicture}
\caption{\textbf{Top panel:} Fraction of stars remaining in stable
  orbits with initial velocities close to $v_T=0\kms$ for seven
  simulations in the range $v_{\odot,\mathrm{reflex}}=-220$ to
  $-260\kms$. \textbf{Middle panel:} Same as top panel, but for $|z|<1$
  kpc. \textbf{Bottom panel:} Same as top panel, but for $|z|<50$
  pc.}
\label{fig:dips}
\end{figure}
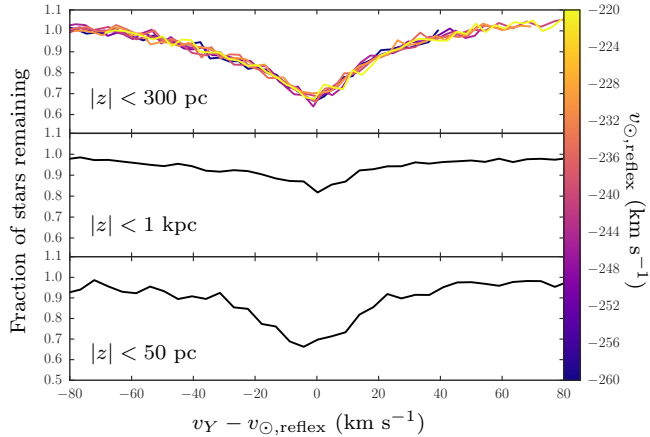

Figure~\ref{fig:dips} shows the computed dip profiles for velocities
within $80\kms$ of $v_T=0$. The top panel shows the results of 7
simulations with reflex solar motion, $v_{\odot,\rm{reflex}}$, of
$-220$, $-230$, $-235$, $-240$, $-245$, $-250$ and $-260\kms$,
assuming $|z|<300$ pc as the criterium `close to the disc plane'. It
is clear that the shape and depth of the overlaid distributions are
very similar and thus not dependent on the value of the solar
motion. The dip profile is not entirely symmetric around zero
$v_Y-v_{\odot,\rm{reflex}}$, because the spatial distribution of the
TGAS+RAVE sample is asymmetric around the Sun. The middle panel of
Figure~\ref{fig:dips} displays the same as the upper panel, for
$v_{\odot,\rm{reflex}}=-240\kms$, but assuming $|z|<1$ kpc to weight
the orbits. The profile of the dip is notably shallower, with a depth
of $\sim0.85$, compared to $\sim0.7$ for the models in the top panel,
but the shape of the dip profile is similar. The bottom panel of
Figure~\ref{fig:dips} shows same as the middle panel, but assuming
$|z|<50$ pc to weight the orbits. The depth is consistent with the
models in the top panel.

We create a function $D(v_Y-v_{\odot,\rm{reflex}})$ of the dip by
smoothly interpolating the model with $v_{\odot,\rm{reflex}}=-240\kms$
and $|z|<300$ pc displayed in the top panel of
Figure~\ref{fig:dips}. Because the dip profile computed above is an
approximation, we give the model more freedom by allowing the dip's
amplitude to vary. We model the full distribution $f(v_Y)$ of $v_Y$ as
an exponential multiplied by the dip:
\begin{equation}\label{eq:fit}
\begin{split}
f(&v_Y) \propto\\
& \exp(m\,v_Y+b)\times\left(1-\alpha\,\left[\frac{1-D(v_Y-v_{\odot,\rm{reflex}})}{1-D(0)}\right]\right).
\end{split}
\end{equation}
Written in this manner, the depth is parameterized by a parameter
$\alpha$, normalized such that $\alpha=0$ corresponds to no dip and
$\alpha=1$ corresponds to a dip reaching all the way to zero, i.e., a
complete absence of stars.

\section{Detection of a zero-angular-momentum feature in \emph{Gaia}-DR1}\label{sec:detection}

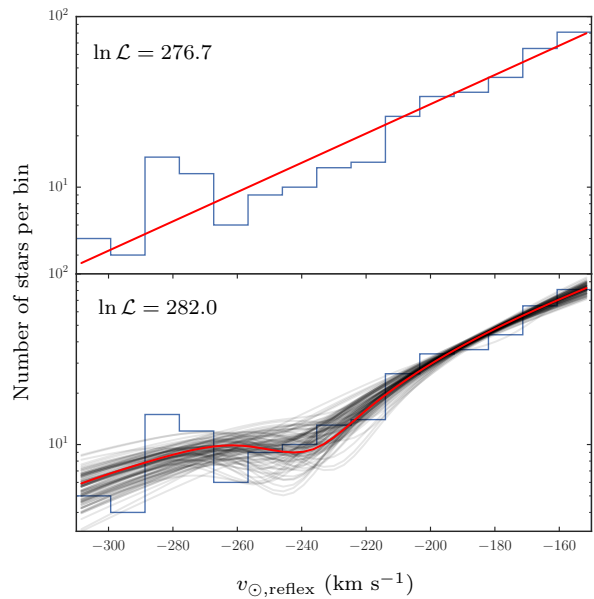
\begin{figure}
\begin{tikzpicture}
  \node (img)  {\includegraphics[width=0.49\textwidth]{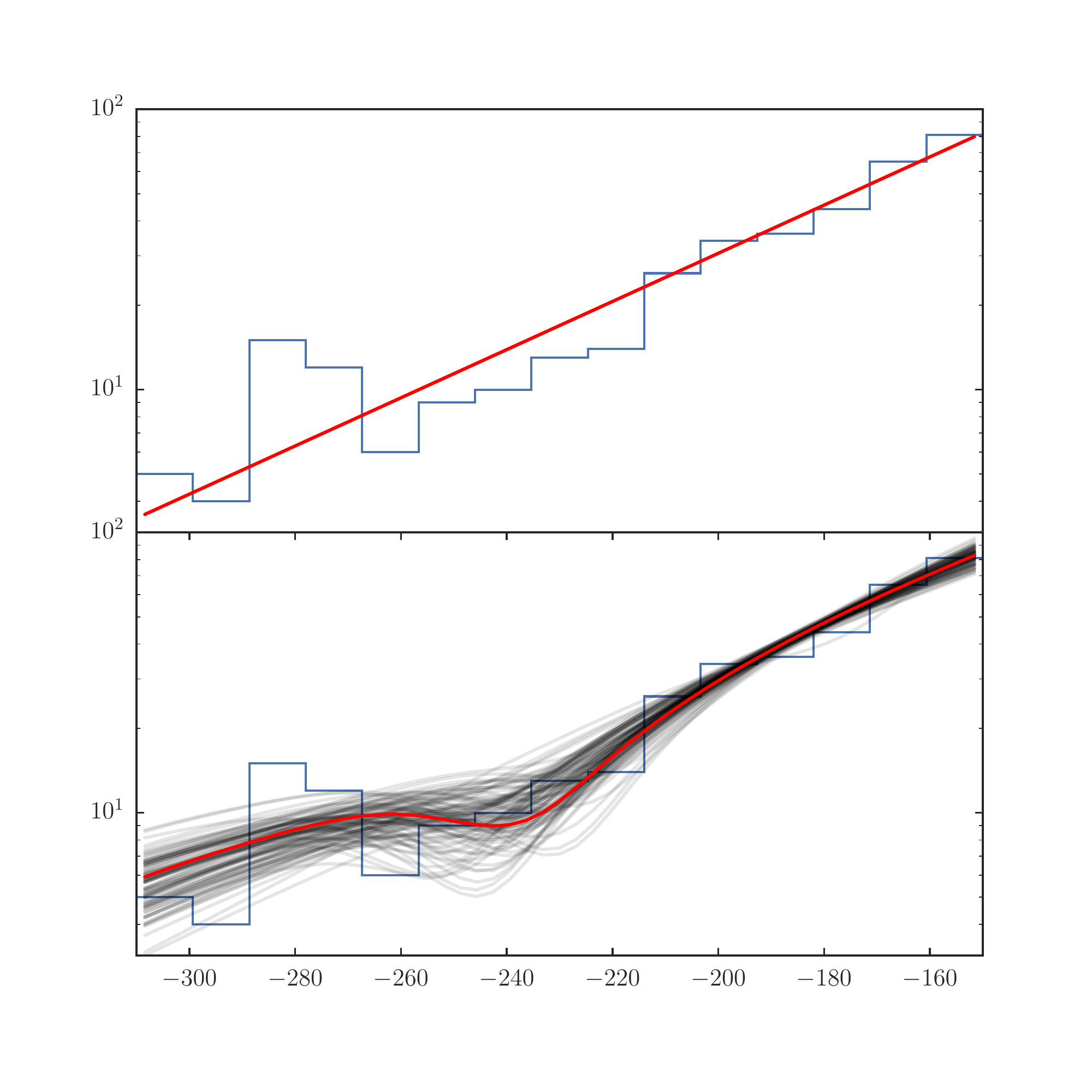}};
  \node[below=of img, node distance=0cm, yshift=8.8cm,xshift=-2.3cm,font=\color{black}] {$\ln\mathcal{L}=276.7$};
  \node[below=of img, node distance=0cm, yshift=5.4cm,xshift=-2.23cm,font=\color{black}] {$\ln\mathcal{L}=282.0$};
  \node[left=of img, node distance=0cm, rotate=90, anchor=center,yshift=-1.5cm,,font=\color{black}] {Number of stars per bin};
\node[below=of img, node distance=0cm, anchor=center,yshift=1.5cm,,font=\color{black}] {$v_{\odot,\mathrm{reflex}}$ (km s$^{-1})$};
\end{tikzpicture}
\caption{\textbf{Top panel:} Tail of the $v_Y$ distribution overlaid
  with the best fit model assuming $\alpha=0$. \textbf{Bottom panel:}
  Tail of the $v_Y$ distribution overlaid with the best fit model
  (red) allowing the dip depth $\alpha$ to be a free parameter. We
  also display 100 models randomly sampled for the parameters from
  MCMC. Each panel includes the ln likelihood of the best-fit. The dip
  is detected at $2.1\sigma$ and centered on $-239\kms$.}
\label{fig:fit}
\end{figure}

We fit the distribution of $v_Y$ of disk stars ($|v_X| < 150\kms$ and
$|v_Z| < 150\kms$) in the range $-310\kms \leq v_Y < -150\kms$ using
the model in Equation~\eqref{eq:fit}. The top panel of
Figure~\ref{fig:fit} displays the best fit $\alpha=0$ model overlaid
on the zoomed in tail of the histogram of the distribution of
$v_Y$. This clearly demonstrates the dearth of stars between
approximately $v_Y=-210\kms$ and $-270\kms$ when compared with the
exponential model.

The best-fitting dip model is shown in the bottom panel of
Figure~\ref{fig:fit}. The data prefer a dip: the log likelihood of the
best-fit model is $\ln \mathcal{L}=282.0$, compared with $\ln
\mathcal{L}=276.7$ for the best-fit exponential model with two fewer
parameters ($\alpha =0$). For comparison, we also fit a model where we
allow the width to be controlled by a new parameter $w$ that stretches
the profile in Figure~\ref{fig:dips} along the $x$ axis. For this
model we find $\ln \mathcal{L}=282.7$ and $w=1.8$. Although the
likelihood is marginally higher for this model, the significance of
the detection is lower owing to the extra free parameter.

Using the Akaike information criterion \citep[AIC,][]{AIC} the difference in ln likelihood corresponds to a $2.1\sigma$ detection for the $w=1$ model and a $1.8\sigma$ detection for the $w$ free case. When combined with the $1.7\sigma$ detection from the $v_X$ to $v_Y$ analysis using a Bonferroni correction to account for the multiple tests \citep[e.g.][]{Dunn} this gives a $2.7\sigma$ detection for $w=1$ and a $2.5\sigma$ detection when $w$ is left free. We explore the allowed range of models for $w=1$ using a Markov Chain Monte Carlo (MCMC) analysis with \texttt{emcee} \citep{Foreman13a}. We find that $\alpha=0.52\pm0.1$ and $v_{\odot,\rm{reflex}}=-239\pm9\kms$ at $1\sigma$.

Even though we do not have an ab-initio prediction of the dip profile,
we find that the best-fitting $\alpha$ is consistent within its
uncertainty to that obtained using the models (which have $\alpha
\approx 0.3$). Thus, systematics due to our approximate model for the
dip in a simple axisymmetric Milky-Way potential are likely quite
small.

\section{Discussion and outlook}\label{sec:conclusion}

We have shown that a dip in the low-velocity tail of stars in the
solar neighborhood is present at $2.7\sigma$ significance. This feature
can be plausibly explained by the absence of stars on orbits with
approximately zero angular momentum, due to such stars being scattered
onto halo orbits after interaction with the Galactic nucleus as
predicted by \citet{CI87}. Using orbit integration in a realistic
model of the Galactic potential, we have demonstrated that this dip
should be centered on minus the Solar rotational velocity
$v_{\odot,\rm{reflex}} = -v_\odot$. In this way, we were able to
measure $v_\odot=239\pm9\kms$. Our modeling does not include the
Galactic bar or spiral arms, which affect the orbits of
stars. However, because the dip is due to zero-angular momentum stars
moving on orbits that are highly eccentric that are being lost from
the solar neighborhood over many dynamical times, the gravitational
perturbations and current position of the relatively diffuse stellar
bar and of the spiral structure should not greatly affect the profile
of the dip and we expect their effect to be minor on our derived value
of $v_{\odot,\rm{reflex}}$.

Observations of the Galactic center have precisely measured the proper
motion $\mu_{\mathrm{Sgr\ A}^*}$ of Sgr A$^*$
\citep{Reid04a}. Assuming that Sgr A$^*$ is at rest with respect to
the dynamical center of the Milky Way, this apparent proper motion is
due to the reflex motion of the Sun. The ratio of the linear and
angular measurements of the reflex motion of the Sun
$v_{\odot,\rm{reflex}}/\mu_{\mathrm{Sgr\ A}^*}$ then becomes a
measurement of the distance $R_0$ to the Galactic centre. For
$\mu_{\mathrm{Sgr\ A}^*} = 30.24\pm0.11\kms\kpc\inv$, our measurement
of $v_{\odot,\rm{reflex}}$ implies $R_0 =
7.9\pm0.3\kpc$. This is an essentially dynamical measurement of $R_0$. If the
lack of zero angular momentum stars is confirmed in future \emph{Gaia}
data releases and systematics can be controlled, this feature can in
principle saturate the precision implied by the uncertainty in
$\mu_{\mathrm{Sgr\ A}^*}$, which is $\approx30\pc$.

Future $Gaia$ data releases will have many orders of magnitude more
stars, with substantially higher accuracy on the measured
parameters. Additionally, future $Gaia$ data releases will contain
line-of-sight velocities for stars in the Solar neighbourhood,
removing the asymmetry in the sample introduce by the necessity to
cross match with RAVE. Thus, with future $Gaia$ data releases we hope
that this technique may provide a very precise measurement of the
solar reflex motion with respect to the Galactic centre. We can make a
rough prediction for the strength of the feature expected in future $Gaia$ data releases by using the $Gaia$ Object Generator \citep[GOG,][]{Luri} based on the $Gaia$ Universe Model Snapshot
\citep[GUMS,][]{GUMS}. GOG predicts that approximately 17.5 million stars will be observed with radial velocities within 700 pc, and thus, by weighting that number by the fraction of
stars in the tail compared to the full sample matched with RAVE,
e.g. 374/216,201=0.0017, we can predict that $Gaia$ will observe roughly
30,000 stars in the tail within 700 pc. This should allow us to confirm the existence of this feature, and if present allow us to increase the precision of the measurement of $v_{\odot,\rm{reflex}}$ to within approximately $1\kms$, assuming the errors decrease on the order of $\sqrt{N}$. In turn the error on $R_0$ would become dominated by the error in the proper motion measurement. At this level the systematics will then become extremely important, and in future analysis we must take into consideration other factors such as the bar, spiral structure and local giant molecular clouds, although the tangential deflections are expected to be small.

Further work is also needed to fully determine the origin of this
feature, and to make more detailed predictions on the shape and depth
which we expect to observe. The exact dimensions of the feature are
likely slightly dependent on the potential of the inner Galaxy, and
thus matching the observed shape in future $Gaia$ data releases to dip
functions constructed from various models for the potential may give
us valuable insight into the potential of the inner Galaxy.

\acknowledgements J.A.S.H. is supported by a Dunlap Fellowship at the
Dunlap Institute for Astronomy \& Astrophysics, funded through an
endowment established by the Dunlap family and the University of
Toronto. J.B. and R.G.C. received partial support from the Natural
Sciences and Engineering Research Council of Canada. J.B. also
received partial support from an Alfred P. Sloan Fellowship. The MCMC
analyses in this work were run using \emph{emcee} \citep{Foreman13a}. 

This work has made use of data from the European Space Agency (ESA)
mission {\it Gaia} (\url{http://www.cosmos.esa.int/gaia}), processed
by the {\it Gaia} Data Processing and Analysis Consortium (DPAC,
\url{http://www.cosmos.esa.int/web/gaia/dpac/consortium}). Funding for
the DPAC has been provided by national institutions, in particular the
institutions participating in the {\it Gaia} Multilateral
Agreement. Funding for RAVE (\url{http://www.rave-survey.org}) has been provided by
institutions of the RAVE participants and by their national funding
agencies.

\end{document}